\documentclass[journal=jacsat,manuscript=article,layout=twocolumn]{achemso}

\newcommand*{\sometext}{The interaction of silicene with Si, C, H, O, Ti atoms along with H$_2$, H$_2$O and O$_2$ molecules are investigated and the induced functionalities thereof are analyzed using first principles density functional theory. Si adatom initially adsorbed at the top site of silicene pushes down the Si atom underneath to form a dumbbell like structure with 3+1 coordination. This prediction is important for silicene research and reveal new physical phenomena related with the formation of multilayer Si, which is apparently the precursor state for missing layered structure of silicon. We found that dumbbell structure attributes coverage dependent electronic and magnetic properties to nonmagnetic bare silicene. Even more interesting is that silicene with dumbbells is energetically more favorable than the pristine silicene: The more dense the dumbbell coverage, the stronger is the cohesion. Incidentally, these structures appear to be intermediate between between silicene and silicon. Carbon adatom, which is initially adsorbed to the bridge position, substitutes one Si atom, if it overcomes a small energy barrier. Oxygen molecule can dissociate on silicene surface, whereby constituent oxygen atoms oxidize silicene by forming strong bonds. By varying the concentration and decoration of carbon, hydrogen and oxygen atoms one can tune the band gap of silicene. Through the adsorption of hydrogen or titanium adatom, silicene acquires spin polarized state. A half metallic ferromagnetic behavior is attained at specific uniform coverage of Ti adatom, which may function as a spin valve.

{\bf Keywords:} dumbbell structure, silicon oxide, silicon carbide, multilayer silicon}

\let\oldmaketitle\maketitle
\let\maketitle\relax

\author{V. Ongun \" Oz\c celik}
\affiliation{UNAM-National Nanotechnology Research Center, Bilkent University, 06800 Ankara, Turkey}
\alsoaffiliation{Institute of Materials Science and Nanotechnology, Bilkent University, Ankara 06800, Turkey}
\email{ongunozcelik@bilkent.edu.tr}
\author{S. Ciraci}
\affiliation{UNAM-National Nanotechnology Research Center, Bilkent University, 06800 Ankara, Turkey}
\alsoaffiliation{Institute of Materials Science and Nanotechnology, Bilkent University, Ankara 06800, Turkey}
\alsoaffiliation{Department of Physics, Bilkent University, Ankara 06800, Turkey}
\email{ciraci@fen.bilkent.edu.tr}

\title{Local Reconstructions of Silicene Induced by Adatoms}

\begin{document}

\twocolumn[
\begin{@twocolumnfalse}
\oldmaketitle
\begin{abstract}
\sometext
\end{abstract}
\end{@twocolumnfalse}
]

\section{Introduction}
Just after the synthesis of quasi 1D tubular\cite{iijima1,iijima2} and 2D single layer graphene\cite{novo1,novo2} structures, which have exhibited exceptional properties, answers to the critical question of whether Si can form similar structures have been desperately sought.\cite{takeda,zhang,engin} This question was rather logical, since C and Si are group IV elements of the periodic table and hence they are isovalent. Also it has been  attempting to adopt the technology developed behind the Si crystal to Si nanostructures. Unfortunately, the attempts of searching single layer Si in honeycomb structure have been discouraged usually by the arguments that Si does not have layered allotrope like graphite. Surprisingly, free standing silicene and germanene, namely graphene like single layer honeycomb structures of silicon and germanium, have been shown to be stable.\cite{seymur2009} Also it has been shown that silicene shares several of the exceptional properties of graphene, such as the linearly crossing of $\pi$- and $\pi^*$-bands at the Fermi level (if a very small gap opening of $\sim$ 1 meV due to the spin-orbit coupling is neglected),\cite{liu} the ambipolar effect and the family behavior observed in nanoribbons.\cite{seymur2010} Although the strong $\pi - \pi$ coupling ensures planar geometry of graphene, this coupling weakens in silicene. However, the endangered stability is regained by the rehybridization of $3s$ and $3p$ valence orbitals to four-fold $sp^3$-like bonds through the dehybridization of three-fold planar $sp^2$ bonds. This leads to the buckling of the planar honeycomb structure. Accordingly, single atomic plane of graphene is replaced by two atomic planes, which are split by a buckling of $\Delta$=0.44\AA~ and the alternating Si atoms at the corners of the hexagons are located in different atomic planes. Owing to the buckling (or puckering) of the structure, the vertical symmetry can be broken under perpendicular electric field which polarizes the Si atoms and hence opens a band gap.\cite{ongunvacuum, falko, liang2012, ezawa2012} Presently, silicene has been an active field of research with several challenges.\cite{lebeg,garcia,bai,pan,wang,angel} Search on the similar single layer, honeycomb topology have extended to group IV-IV, III-V and II-VI compounds\cite{engin,jiang,akturk,hasan1} including SiC\cite{sic} and ZnO,\cite{kulkarni,zno} as well as transition metal oxides and dichalcogenides.\cite{can1,can2} Each of these structures has been predicted to display interesting properties with potential applications in nanotechnology. Nowadays, great expectations for fundamental properties and technological application have been directed towards various single layer nanostructures and their composite, van der Waals thin films.\cite{search}

Our earlier predictions\cite{engin,seymur2009,seymur2010} on the stability of silicene and its properties have been confirmed recently by realizing the growth and characterization of single layer silicene on Ag substrate.\cite{lay2010, lay2012} These studies, at the same time, initiated a growing interest in silicene. Even though two dimensional semi-metallic silicene has limited applications for nanoelectronics, several new functionalities can be achieved by chemisorption of a number of foreign atoms or molecules.\cite{peter1,peter2, lin2012}

Motivated by the self-organized behavior of carbon host atom on graphene,\cite{cac1,cac2,canethem} here we investigate the effects of silicon and carbon adatom on silicene. Silicon adatom is a host atom and its interaction with silicene may be crucial for the study of the growth of the multilayer silicene. Additionally, the decoration through uniform Si coverage may attribute useful functionalities to silicene. Being a group IV element, carbon adatom is isovalent to Si and its interaction with silicene is also important for future graphene-silicene composite materials like silicene including SiC quantum dots. Apart from Si and C adatoms, we also study the interactions of silicene with of H, O, Ti adatoms and H$_2$, H$_2$O and O$_2$ molecules. Oxygen and hydrogen atoms are known to form strong bonds on the surface of bulk silicon and are crucial elements in Si based microelectronics. In particular, it is important to know whether silicene is oxidized as easily as bulk Si does. Finally, Ti is a $3d$-transition metal atom and may form strong bonds with silicene for metallic contacts like Ti on graphene. Therefore, the interaction of these atoms with silicene and the selected functionalities attained thereof have been the focus of this work. Our study is based on free standing (suspended) silicene, since it has been shown to be stable\cite{seymur2009} like graphene. Even if silicene is grown on Ag(111) and hence its states engage in hybridization with those of Ag(111) inducing some changes in specific states, the effects of hybridization cease at the top surface of multiple silicene layers. In this respect, our study is relevant for the strong interaction between Si, C, O, H and Ti adatoms and silicene surface.

We show that the interactions between Si adatom and silicene are complex and lead to amazing results, which are of crucial importance in the rapidly developing research on silicene: Silicon adatom on silicene pushes the Si atom underneath and readily forms a dumbbell (D) structure by donating significant electronic charge to nearest Si atoms. By engaging in a $3+1$ coordination,  D decorated silicene is a structure between the fourfold coordinated diamond structure and the single layer, buckled silicene. Hence, silicene with dumbbells is slightly more favorable than pristine silicene. The cohesion of uniform D+silicene structure becomes even more superior to that of silicene when the smallest D-D distance, $d_{D-D}$, is less than $\sqrt{3} a$ ($a$ being the lattice constant of pristine silicene). Dumbbells also display interesting dynamics and structural transformations, which are crucial for the understanding of the growth of multilayer silicene. Additionally, we also present the coverage dependent features of the D+silicene structure.

Carbon adatom also creates unique reconstructions in silicene. Carbon initially forms a bond on the bridge site; however if a small barrier is passed, the C adatom substitutes one of the host silicene atoms to form a substitutional impurity. On the other hand, oxygen molecule can dissociate on silicene, where constituent oxygen atoms form strong bonds to oxidize silicene. A bridge bonded O adatom can pass from the top side to the bottom side of silicene once a small energy barrier is overcame and thus it can easily penetrate across silicene layers. H and Ti adatoms attribute magnetic properties. Silicene acquires integer magnetic moment and half-metallic character due to a specific uniform coverage of Ti.

\begin{table*}
\caption{Characterization of the case, where a single adatom (Si, C, H, O and Ti) adsorbed uniformly to each (4x4) supercell of silicene (corresponding to the coverage $\Theta$=1/32). $BS$: Binding site, where D, S, T, B and TH represents dumbbell structure, substitution of adatom, top site, bridge site and between top and hollow sites, respectively; $d$: The smallest distance between the adatom and the nearest Si atom; $E_b$: Binding energy in eV per unit cell and in kJ per mol; $E_f$: Formation energy; $E_B$: The minimum energy barrier in the migration of adatom; $E_g$: The smallest band gap between spin-unpolarized conduction and valance bands. For spin-polarized systems the gap between spin up and spin down/spin up and spin up /spin down and spin down conduction and valence bands are given; $Q^*$: Effective charge on the adatom; $\mu$: the total magnetic moment per supercell. *Note that when 4 Ti adatoms are adsorbed on (4x4) supercell of silicene uniformly, the resulting structure is a half-metal (HM) with ferromagnetic order.}
\label{table1}
\begin{center}
\begin{tabular}{ccccccp{4cm}cc}
\hline  \hline
Adatom & $BS$ & $d$ (\AA) & $E_b$ (eV, kJ/mol) & $E_f$ (eV) & $E_B$ (eV) & $E_g$ ($\uparrow \downarrow / \uparrow \uparrow / \downarrow \downarrow$)(eV)  & $Q^*$ (e) & $\mu (\mu_B$)  \\
\hline
Si & D & 2.39 & 3.96 (380) & -0.75 & 0.92 & 0.08/0.43/0.43 & 0.22 & 2.0 \\
C & S & 1.85 &  5.88 (564) & -1.77 & 1.52 & 0.19 & -0.42 & 0 \\
H & T & 1.51 &  2.12 (203) & -0.13 & 0.26 & 0.16/0.29/0.24 & -0.20 & 1.0 \\
O & B & 1.70 &  6.16 (591) & 2.83 & 0.65 & 0.21 & -0.34 & 0 \\
Ti & TH & 2.50 &  4.14 (397) & -3.59 & 0.22 & 0.09/0.09/0.15 (HM*) & 0.16 & 2.0 \\
\hline
\hline
\end{tabular}
\end{center}
\end{table*}

\section{Method}
We have performed spin polarized density functional theory calculations within generalized gradient approximation(GGA) including van der Waals corrections.\cite{grimme06} We used projector-augmented wave potentials PAW,\cite{blochl94} and the exchange-correlation potential is approximated with Perdew-Burke-Ernzerhof, PBE functional.\cite{pbe}

Using the supercell geometry within the periodic boundary conditions, we considered the adatoms as isolated dopants. In large supercells the adatom-adatom coupling is reduced significantly and conditions of isolated adatom is approximately met. The localized states of dopants appear as flat bands. Very large supercells are not convenient from the computation point of view. Therefore, one has to optimize the supercell size. In the present study, we used $4 \times 4$ supercells of silicene, which corresponds to the uniform coverage of one Si adatom per 32 Si host atom, namely $\Theta$=1/32. The size of this supercell is tested to be sufficient to minimize the adatom-adatom coupling for the purpose of the present study. On the other hand, results obtained from relatively smaller supercells are taken for the uniform coverage or decoration of adatoms. Hence, in specific cases, we also treated a uniform coverage of $\Theta >$ 1/32 to examine the effects of significant couplings between adatoms.

The Brillouin zone was sampled by (11x11x1) \textbf{k}-points in the Monkhorst-Pack scheme where the convergence in energy as a function of the number of \textbf{k}-points was tested. The number of \textbf{k}-points were further increased to (17x17x1) in small supercell calculations.  Atomic positions were optimized using the conjugate gradient method, where the total energy and atomic forces were minimized. The energy convergence value between two consecutive steps was chosen as $10^{-5}$ eV. A maximum force of 0.002 eV/\AA~ was allowed on each atom. Numerical calculations were carried out using the VASP software.\cite{vasp}

\begin{figure*}
\includegraphics[width=16cm]{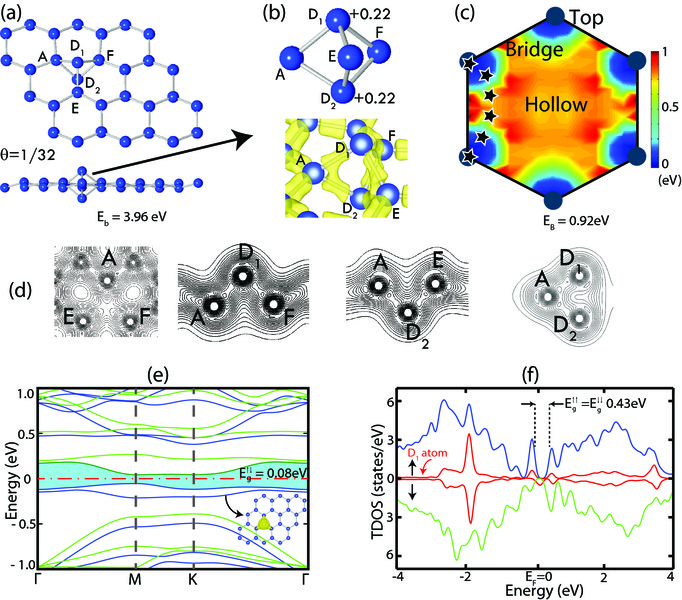}
\caption{(Color online) One Si adatom adsorbed to each (4x4) supercell of silicene, which corresponds to the uniform coverage of $\Theta$=1/32. (a) Top and side views of the atomic configuration of the dumbbell (D) structure. Blue balls represent Si atoms. (b) Magnified view of the D structure together with the isosurface charge density. D$_1$ and D$_2$ denote Si atoms at both ends of the dumbbell; and A, E and F are silicene atoms nearest to D$_1$ and D$_2$. Excess charges on the Si atoms of the dumbbell structure are shown by numerals. (c) Energy landscape for the Si adatom on silicene calculated on a hexagon. The migration path of the Si adatom with minimum energy barrier $E_B$ is indicated by stars. (d) Contour plot of the total charge density $\rho_{T}(\textbf{r})$,  on the horizontal plane passing through A, E and F atoms, and on the planes passing through A-D$_1$, A-D$_2$ and D$_1$ -D$_2$ bonds. (e) Energy band structure of the D+silicene structure with the dash-dotted line indicating the Fermi level. Blue(dark) and green(light) lines represent spin up and spin down states, respectively. The inset shows that the isosurface charge density of spin up states making the flat band just below the Fermi level is localized mainly at the D-structure. (f) Spin projected total density of states TDOS. Up-arrow and down-arrow stand for spin up and spin down states, respectively. The density of states DOS projected to $D_1$ is augmented four times and plotted in panel (f).}
\label{fig1}
\end{figure*}

The binding energy $E_b$, is calculated from the expression, $E_b=E_{T}[silicene]+E_{T}[A/M]-E_{T}[A/M+silicene]$, in terms of the total energies of bare silicene supercell and of free adatom A, (=Si, C, H, O or Ti) or molecule M, (M=H$_2$, O$_2$, H$_2$ or H$_2$O), and the structure optimized total energy of one A or one M adsorbed to each silicene supercell, respectively. All total energies are calculated in the same supercell. In our notation, $E_{b} >$0 indicates a binding structure. The formation energy, $E_f$, takes into account the binding relative to the ground state of A either in bulk crystal or in a molecule. Accordingly, for the case of hydrogen, for example, $E_{f} =-E^{'}_{b}/2+E_b$, where $E^{'}_{b}$ is the binding energy of H$_2$ relative to hydrogen atom. Normally, while a process with $E_{f}>$0 is favored, $E_{f} <$0 may give rise to clustering or desorption under specific circumstances.

The bonding and effective charge of adatoms are characterized by calculating total charge density of adatom+silicene system, $\rho_{T}(\textbf{r})$. We presented the charge distribution in terms of isosurfaces and contour plots. We also carried out Mullliken analysis\cite{mulliken} in terms of atomic orbitals of constituent atoms to obtain their effective charge, $Q^*$, as implemented in the SIESTA package.\cite{siesta} The energy landscape of an adatom on silicene is calculated by placing the adatom  to 500 different grid points in the hexagon of silicene and performing self consistent energy minimization calculation for each point. At each grid point, the $x$ and $y$ coordinates of the adatom were kept fixed while its $z$ coordinate was relaxed to minimize the total energy. The energy at each grid point is designated by a color code. The fully relaxed, spin-polarized calculations of Ti adatoms are also repeated using local basis set in SIESTA package resulting in the same magnetic moment obtained by using plane wave basis set.\cite{vasp}

\section{Results and Discussions}
The interactions of silicene with the adatoms (Si, C, H, O and Ti) and molecules (H$_2$O, H$_2$ and O$_2$) are characterized by the calculation of optimum binding geometries and the corresponding atomic structures at the proximity of the adatom. The energy landscapes and the path of migrations with lowest energy barriers are calculated; relevant electronic and magnetic properties together with the spin-projected and adatom projected densities states are presented. We present a summary of our results in Table~\ref{table1}. Here the top (T), bridge (B), hollow (H), between top and hollow (TH), and between bridge and hollow (BH) are special sites where adatoms are bound to silicene. In specific cases, the adatom can substitute the host Si atom (S) or can push it down to form a dumbbell (D) structure.

\begin{table*}
\caption{Variation of the properties of D+silicene (i.e. silicene with uniform coverage of dumbbell structure in different supercells). ($n$x$n$): Supercell size; $\Theta =1/2n^2$: uniform coverage of D structure per the number of Si atoms in silicene supercell; $(2n^2+1)$: Number of Si atoms of D+silicene in the ($n$x$n$) supercell of silicene; $d_D$: distance between two nearest neighbor dumbbell structures; $\mu$:magnetic moment per supercell;  $ES$: Electronic structure specified as metal M, or semiconductor with the band gap between valance and conduction bands, $E_g$; $E_b$: binding energy; $E_{C}[D]$: cohesive energy of Si atom in D+silicene structure; $\Delta E_{C}$: difference between the cohesive energies of Si atom in D+silicene and pristine silicene, where positive values indicates that D+silicene structures are favorable. The cohesive energy of pristine silicene is $E_C$=3.936 eV.}
\label{table2}
\begin{center}
\begin{tabular}{cccccccccc}
\hline  \hline
Supercell & $\Theta$ & $(2n^2+1)$ & $d_D$ (\AA) & $\mu$ ($\mu_B$) & ES &  $E_b$ (eV) & $E_{C}[D]$ (eV) & $\Delta E_{C}$ (eV) \\
\hline
$1 \times 1$ & 1/2  & 3  & 3.58  & 0    & M         & 4.13 & 4.002 & 0.066 \\
$ \sqrt{3} \times \sqrt{3}$ & 1/6  & 7  & 6.52  & 0    & M         & 4.35 & 4.001 & 0.065 \\
$2 \times 2$ & 1/8  & 9  & 7.70  & 0    & M         & 3.89 & 3.956 & 0.020 \\
$3 \times 3$ & 1/18 & 19 & 11.50 & 1.8  & M         & 3.92 & 3.945 & 0.009 \\
$4 \times 4$ & 1/32 & 33 & 15.40 & 2.0  & 0.083     & 3.96 & 3.939 & 0.003 \\
$5 \times 5$ & 1/50 & 51 & 19.20 & 2.0  & 0.078     & 4.02 & 3.937 & 0.001 \\
$7 \times 7$ & 1/98 & 99 & 23.0  & 2.0  & 0.075     & 4.03 & 3.938 & 0.002 \\
\hline
\hline
\end{tabular}
\end{center}
\end{table*}

\begin{figure*}
\includegraphics[width=16cm]{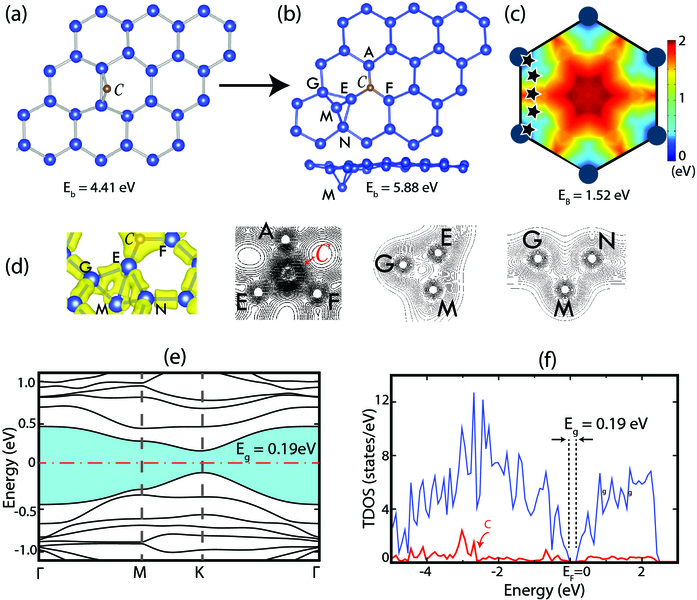}
\caption{(Color online) One carbon adatom adsorbed to each (4x4) supercell of silicene amounting to a uniform coverage of $\Theta$=1/32. Blue and brown balls indicate Si and C atoms, respectively.(a) Top view of atomic configuration of C adatom adsorbed above a Si-Si bond (BH-site) with a binding energy of $E_b$=4.41 eV. This is a precursor state for the substitutional carbon atom described in (b). (b) Top and side views of the carbon atom substituting one of the Si atoms and pushing it to point M to form three fourfold coordinated Si atoms at (G,E,N). The binding energy of the C adatom at this configuration is $E_b$=5.88 eV (564kJ/mol). (c) The energy landscape of the C adatom on silicene calculated on a hexagon. The migration path of the Si adatom with minimum energy barrier $E_B$=1.52 eV is indicated by stars. (d) Isosurfaces of the total charge density, $\rho_{T}(\textbf{r})$ describing the bond formation around the substitutional carbon atom. Contour plots of the total charge density in the plane passing through A, E and F atoms show the $sp^2$ hybrid bonds of the substitutional carbon atom. Contour plots in the plane of G,E,M and G,M,N atoms show four-fold bond formation. (e) Energy band structure of silicene including a single substitutional carbon atom in each (4x4) supercell. (f) The corresponding total density of states TDOS and the density of states DOS projected on substitutional carbon atom.}
\label{fig2}
\end{figure*}

\subsection{Si adatoms}
A Si adatom, which is initially bonded to the T site, pushes down the Si atom underneath to form a dumbbell D structure. This configuration occurs without any barrier and has the binding energy of $E_b$=3.96 eV for a single Si adatom forming a D-structure in each $4 \times 4$ supercell, which is equivalent to 380kJ per one mole of Si atoms. Therefore, a single D structure is not a predetermined configuration; it can occur while Si adatom migrates on silicene. In ~\ref{fig1}(a-b) the atomic configuration of D structure is presented. Two Si atoms positioned at two ends of the dumbbell are specified as D$_1$ and D$_2$. The distance between the dumbbell atoms D$_1$ or D$_2$ and nearest silicene atoms (A, E and F), which are located in a horizontal plane is 2.41 \AA. This is larger than the nearest Si-Si distance 2.28 \AA~ in pristine silicene. The distance between D$_1$ and D$_2$ is relatively large and is 2.69 \AA. We note that in our earlier work, it was found that carbon atom migrating on graphene can form also similar dumbbell structure, even if it is slightly less energetic relative to its B-site binding.\cite{canethem} Recently, formation of the dumbbell structure in silicene layers has been deduced during the course of vacancy healing. \cite{healing}

Our recent calculations have also demonstrated that the formation process of the dumbbell structure on silicene multilayers grown on Ag substrate is practically identical to that on free-standing silicene reported in the present paper. The reconstructions such as 3$\times$3 and $\sqrt{3} \times \sqrt{3}$ that took place in the course of multilayer growth on Ag are induced by the formation of dumbbell structure. Therefore, the predictions in the present paper are important for silicene research and reveal new physical phenomena related with the formation of multilayer Si, which is apparently the precursor state for missing layered structure of silicon.

The Mulliken analysis indicates that the depletion of electronic charge from each of D$_1$ and D$_2$ atoms is +0.22 electrons, which is transferred to nearest three Si atoms of silicene. This situation suggests that strong bonds with mixed covalent-ionic character\cite{harrison} form between nearest silicene atoms (A, E and F) and each of the dumbbell atoms, D$_1$ or D$_2$. On the other hand the D$_1$ - D$_2$ bond is relatively weak. These arguments can be depicted from isosurfaces of the total charge density and the charge density contour plots presented in ~\ref{fig1}(d). Accordingly, each of A, E and F atoms are four-fold coordinated, and hence they mimic the bulk Si crystal by making four bonds with their nearest neighbors. Whereas D$_1$ and D$_2$ atoms are 3+1 coordinated, each of them makes three strong bonds with A, E and F, but are weakly bonded to each other. We note that having positively charged two Si atoms located above and below the Si planes of buckled silicene may attribute interesting functionalities, which may be monitored by the electric field applied perpendicular to silicene. For example, positively charged surface of D+silicene is attracted by negatively charged surfaces or vice versa for positively charged surfaces. Additionally, the work function (or photoelectric threshold) of silicene increases upon its decoration with D.

It should be noted that the formation energies of both the pristine silicene and D+silicene are negative with respect to bulk Si in diamond structure. In spite of that, one free Si adatom at the close proximity of a D structure does not form any bond with D$_1$ or D$_2$ to nucleate a cluster or an atomic chain as carbon adatom does on graphene or boron nitride\cite{cac1, cac2} It rather moves to the third nearest neighbor and form another D structure. It appears that the D structure display a self-organizing character. The D structure occurs at the T sites of silicene; H-sites are unfavorable since Si adatom cannot form sixfold long bonds with Si atoms at the corners of the hexagon. The calculated energy landscape of the Si adatom is shown in ~\ref{fig1}(c). The minimum energy barrier for the migration of Si adatom is estimated to be 0.92 eV. Although this barrier is significant to hinder diffusion at room temperature, at high temperatures the D structure may display interesting dynamics in the course of the growth of silicene.

The fact that the binding energy of the D structure ($E_b$=3.96 eV) is slightly higher than the cohesive energy of a Si atom forming a pristine silicene ($E_C$=3.94 eV) brings about the question whether the D+silicene with diverse decoration of D can be energetically more favorable than bare silicene and may constitute its complex derivatives. To this end, we compare the cohesive energies of Si atoms in the D+silicene structures with diverse coverage values with that in the pristine silicene. The cohesive energy per Si atom in an $n \times n$ supercell comprising one single D structure is obtained from the energy difference between the energy of free Si atom $E_{T}[Si]$ and the total energy of the structure of one D per supercell divided by 2$n^2$+1, namely $E_{C}[D]=E_{T}[Si]-E_{T}[D+silicene]/(2n^2+1)$. Similarly, the cohesive energy of Si atom in a $n \times n$ silicene supercell is $E^{o}_{C}=E_{T}[Si]-E_{T}[silicene]/2n^2$. Then, the positive values of the energy difference, $\Delta E_{C} = E_{C} - E^{o}_{C}$ indicates that D+silicene structure is more favorable. For the sake of comparison, the cohesive energy of single Si atom in bulk silicon is calculated with the same parameters to be 4.71 eV. The cohesive energy and relevant properties of D+silicene are calculated as a function of coverage and presented in Table~\ref{table2}.

The cohesive energy $E_C$ of the D+silicene structure decreases with decreasing coverage. It is larger than the cohesive energy of Si atom in silicene and hence is slightly more favorable energetically than pristine silicene. For $n$=1 ($\Theta$=1/2), the D+silicene structure has nonmagnetic ground state; it is metal and has high cohesive energy.  Similarly, for a single D adsorbed to $\sqrt{3} \times \sqrt{3}$ supercell, which is predicted to be a nonmagnetic metal, $\Delta E_{C}$=65 meV per Si atom is significant. Present results confirm the recent study,\cite{kaltsas} which found that $\sqrt{3} \times \sqrt{3}$ coverage stable and has cohesive energy 48 meV per atom higher than that of pristine silicene. We believe that the difference between the calculated cohesive energies occurs from the van der Waals correction taken into account in the present study. For $n$=2 and $n$=3, $\Delta E_C$ decreases and continues to be nonmagnetic metal. However, for $n$=4, 5 and 7, D+silicene attains spin polarized ground state and achieve $\sim$2 $\mu_{B}$ magnetic moment per supercell. Hence, three of them are spin polarized semiconductor with a band gap  $E^{\uparrow \downarrow}_{g} \sim$ 80 meV. For the case of $n$=4, the flat bands at the edges of valence and conduction bands in ~ \ref{fig1}(e) are derived from orbital states, which are localized at the D structure with also minor contributions from other Si atoms. Similar flat bands due to D structure also occur at -2 eV in the valence band as shown in the spin polarized DOS projected to D atoms presented in ~\ref{fig1}(f).

For $\Theta \leq $ 1/32 the spins are polarized and metallic states change into semiconductors. Also, $\Delta E_{C}$ is reduced and becomes smaller than the accuracy limits of present calculations. Apparently, various structures of D+silicene can be considered as the allotropes of the pristine silicene and display variations in the physical properties as a function of the coverage. The D structures forming uniform (1x1), ($\sqrt{3}$x$\sqrt{3}$), (2x2), (4x4), (5x5) and (7x7) supercells form centered hexagons of different sizes on silicene. On the other hand, two D structures contained in the ($\sqrt{3}$x$\sqrt{3}$) and ($n$x$n$) supercells with $n$=3,6,9,.. can form regular honeycomb structure and yield linearly crossing bands.

Finally, the question of whether the dumbbell Si atoms are active sites of silicene or not is investigated through the adsorption of Ti and H$_2$O to the dumbbell Si atoms. We found that similar to bare silicene H$_2$O did not form bonds with D$_1$ adatom. The increase of the binding energy relative to that on bare silicene was only 130 meV. The binding energy of Ti atom to D$_1$ was almost half of the binding energy of Ti atom to bare silicene. We therefore arrive at the conclusion that the D structure of Si adatom gives rise to interesting electronic and magnetic properties, but it does not involve in active chemical reactions that are significantly different from bare silicene.

\begin{figure*}
\includegraphics[width=16cm]{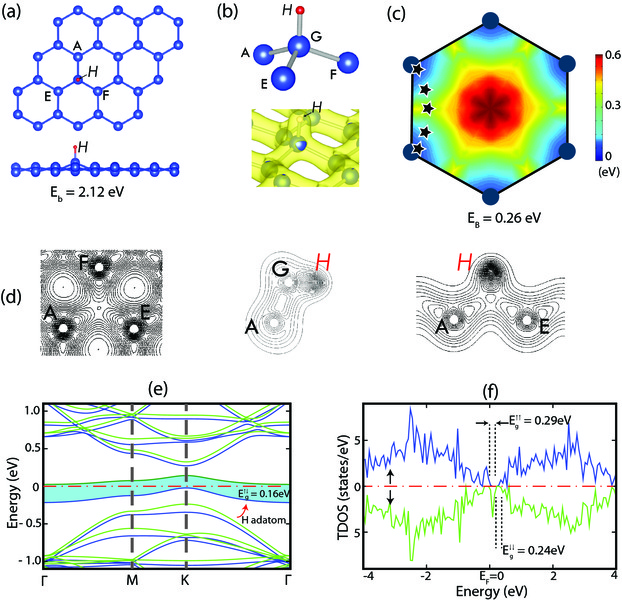}
\caption{(Color online) One hydrogen adatom adsorbed to each (4x4) supercell of silicene. (a) Top and side views of the H adatom adsorbed at the T-site above the protruded Si atom. Blue and red balls represent Si and H atoms, respectively. (b) The local and magnified view of the adsorption geometry with the isosurfaces of the charge density showing bonding configuration. (c) Energy landscape. The path of migration of H with minimum energy barrier of $E_B$=0.26 eV is shown by stars. (d) The contour plots of the total charge density $\rho_{T}(\textbf{r})$ in a lateral plane passing through three Si atoms (A, E, F) nearest to H adatom. Similar contour plots passing through the plane of atoms A, G, H  and A, H, E describe the bonding with the H adatom. (e) Energy band structure of silicene including a single H adatom adsorbed to each (4x4) supercell. Blue(dark) and green(light) lines indicate the spin up and spin down split bands. The flat band of H atom is indicated. (f) The corresponding spin polarized total density states.}
\label{fig3}
\end{figure*}

\subsection{C adatoms}
Free carbon atom is first adsorbed to the BH site with a binding energy $E_b$=4.41 eV. This is only a local minimum and a precursor of another configuration with a higher binding energy of $E_b$=5.88 eV (588kJ/mol), where C adatom substitutes one of the Si atoms of silicene and displaces it to a nearby site below the silicene layer. The latter substitutional site is 1.47 eV more energetic relative to the BH site adsorption, but needs to overcome a small energy barrier. These two configurations are shown in ~\ref{fig2}(a-b). The contour plots of the charge density show dramatic changes in bonding. In particular, substitutional carbon atom acquires 0.42 electrons from three nearest Si atoms (A, E and F), which are not buckled anymore, but locally flattened. Under these circumstances, $sp^3$ like hybrid orbitals are dehybridized upon substitution of carbon atoms, which attains the $sp^2$ bonding with nearest three Si atoms as shown in ~\ref{fig2}(d). We note also that silicene substituted by C atoms can be the precursor of the single layer SiC in honeycomb structure, which is planar and has in-plane stiffness of 166 J/m$^2$.\cite{sic} Additionally, single layer SiC is a wide band gap semiconductor with $E_g$=3.8 eV. Also the Si atom, which is displaced by by substitutional carbon is a potential candidate to form a D structure. It is therefore anticipated that carbon adatom can initiate the D structures since they at the same time grow the planar domains of Si-C compounds in silicene.

The energy landscape presented in ~\ref{fig2}(c) shows that the substitution of C is the most energetic configuration. The minimum energy barrier $E_B$=1.52 eV revealed from this plot is rather high and blocks the migration of C adatom. The migration path of C with lowest energy barrier is marked by stars.

While single substitutional C atom in each (4x4) supercell opens a gap of 0.19 eV as shown in ~\ref{fig2}(e-f), the band gap shall increase with increasing concentration of substitutional C and eventually saturate at $E_g$=3.8 eV. Present findings implies that one can make a mesh of composite material from bare semimetallic silicene by forming C-doped semiconducting domains with tunable band gap.\cite{hasan1} Upon increased C coverage one can also expect to fabricate a core-shell structure consisting of the planar SiC domains and hence quantum dots in semimetallic silicene lattice.\cite{coreshell}

\begin{figure}
\includegraphics[width=8cm]{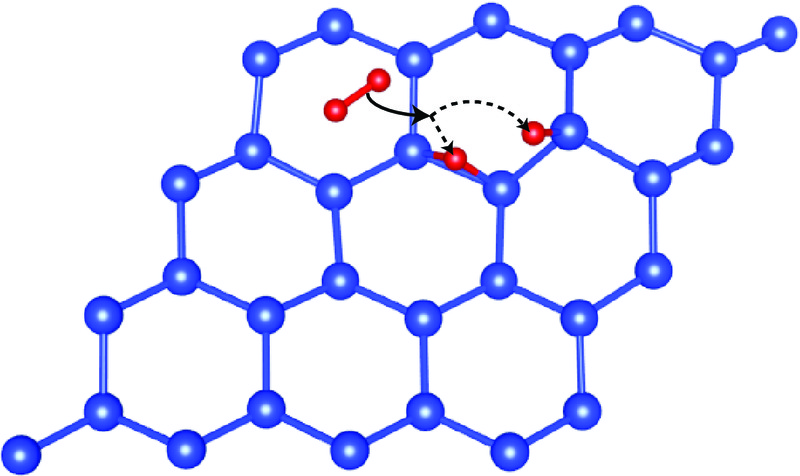}
\caption{(Color online) Dissociation of oxygen molecule on silicene. Dissociated oxygen atoms are adsorbed to different sites on silicene. Blue and red balls represent Si and O atoms, respectively. }
\label{fig4}
\end{figure}

\subsection{Binding of H$_2$ molecule and H atom}
Adsorption of H atom to the surfaces of bulk Si crystal forming monohydride, dihydride and trihydride have been the subject of extensive studies earlier.\cite{sih} The interaction of H$_2$ molecule and H atom is also essential for silicene. Similar to graphane CH\cite{chribbon}, H atom is adsorbed to Si atoms at the corners of hexagons alternatingly from top and bottom surfaces forming silicane i.e. fully hydrogenated silicene, SiH.\cite{sihsil1,sihsil2} Whereas, the interaction between H$_2$ and silicene is rather weak. The dissociation energy of free H$_2$ molecule is 4.5 eV and hence the formation energy for the adsorption of H atom through the dissociation of H$_2$ is negative. Therefore, molecular hydrogen neither forms strong chemical bonds with silicene, nor it dissociates. However, once the atomic hydrogen positioned on the surface of silicene, it can readily form strong bonds with silicene at T-site above the protruded of Si atoms. The binding energy is calculated to be $E_b$=2.12 eV (203kJ/mol). The Si-H bond distance is 1.51 \AA. In ~\ref{fig3} we present all the relevant results related to the interaction and binding of H with silicene.

The isosurfaces of the total charge density shows the bond formation among H adatom and host Si atoms in ~\ref{fig3}(b). As depicted by the contour plots of $\rho_{T}(\textbf{r})$ in ~\ref{fig3}(d), the Mulliken analysis\cite{mulliken} yields an excess charge of $Q^*$=-0.20 electrons on H atom. This situation is in contrast with H adatom on graphene, where charge is transferred from H to graphene, but is in compliance with the ordering of electronegativities\cite{electrone} of Si, H and C atoms as 1.8, 2.1 and 2.5, respectively.

The energy landscape of H adatom shown in ~\ref{fig3}(c) confirms that the T-site is really the energetically most favorable site. We also predict the minimum energy barrier is $E_B$=0.26 eV for the migration of H adatom on silicene. It appears that H adatoms are rather mobile on silicene. In this context, the possibility that the formation of H$_2$ molecule from adsorbed H atoms of SiH leading to the dissociation of H requires serious investigations in studies dealing with silicane.

Hydrogen adatom has dramatic effects on the electronic structure of silicene as shown in ~\ref{fig3}(e-f). First of all, H adsorbed silicene has spin-polarized state, where the spin degeneracy is broken and spin up and spin down bands split. Spin polarization is depicted by the energy band structure of one H atom adsorbed to each (4x4) supercell and corresponding spin projected TDOS. The flat spin up and spin down bands below and above the Fermi level mainly derive from H orbitals. The dispersion of this band decreases with decreasing coverage of H and eventually appears as two localized impurity levels. While the band gap between the highest occupied spin up band and the lowest unoccupied spin down band is only $E_{g}^{\uparrow \downarrow}$=0.16 eV, the gap between spin up bands is $E_{g}^{\uparrow \uparrow}$=0.29 eV.  For this spin polarized ground state of silicene+H adatom system, each H adatom contributes a magnetic moment of $\mu$=1.0 $\mu_B$ per supercell. On the other hand, two H adatom adsorbed to two adjacent Si atoms from different sides of silicene plane has $\mu$=0 $\mu_B$. We conclude this section by noting that the holes created on silicane can have magnetic moments which depend on their size and geometry.\cite{chribbon}

\begin{figure*}
\includegraphics[width=15cm]{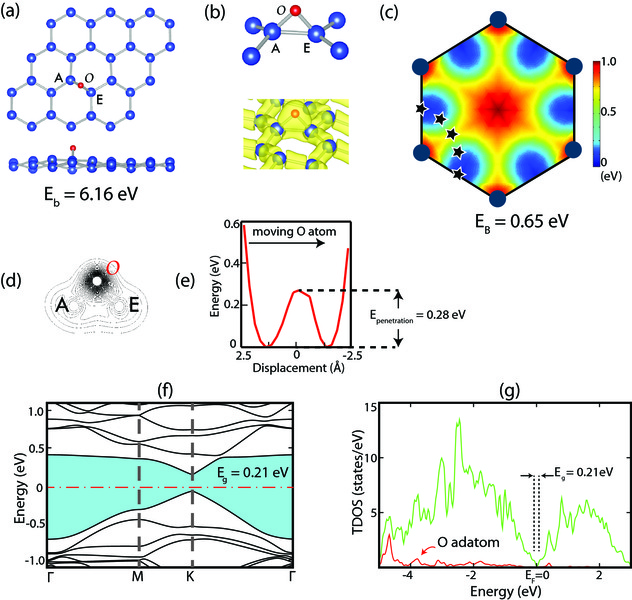}
\caption{(Color online) A single oxygen adatom adsorbed to (4x4) supercell of silicene. (a) Top and side views of the atomic configurations of the O adatom adsorbed at the B-site of silicene above the Si-Si bond. Blue and red balls represent Si and O atoms, respectively. (b) The local and magnified view of the adsorption site with the isosurfaces of the total charge density showing the bonding configuration. (c) Energy landscape. The migration path of O with a minimum energy barrier of $E_B$=0.65 eV is shown by stars. (d) The contour plot of the total charge density $\rho_{T}(\textbf{r})$ in a plane passing through through Si-Si bond and O atom. Weakening of Si-Si bond underneath and charge accumulation on O atoms is clearly seen. (e) Energy variation of the O adatom while it penetrates from upper side to the lower side of silicene. The energy barrier for this process is only 0.28 eV. In the graph, zero in the x-axis indicates the position of the silicene layer. (f) Energy band structure of silicene including the O adatom adsorbed to each (4x4) supercell. The direct band gap at $K$-symmetry point is $E_g$=0.21 eV. (g) The corresponding spin polarized total density states and the density of states DOS projected to O atom.}
\label{fig5}
\end{figure*}

\begin{figure*}
\includegraphics[width=16cm]{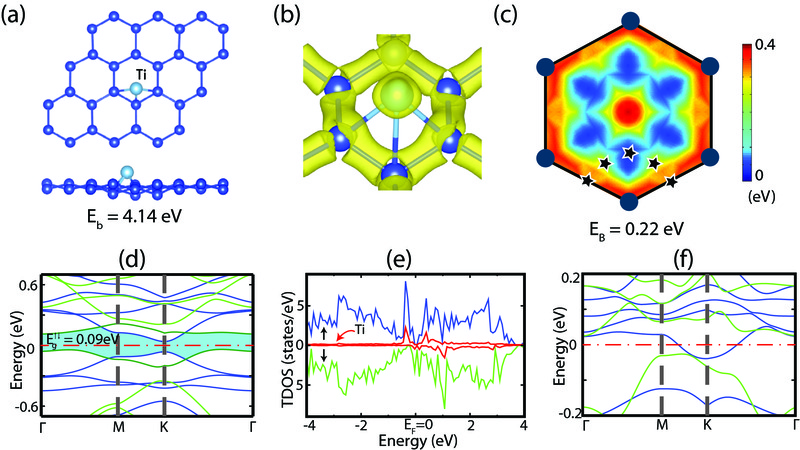}
\caption{(Color online) One Ti adatom adsorbed to each (4x4) supercell of silicene. (a) Top and side view of atomic configuration of Ti adatom adsorbed at the TH site. Dark blue and light blue balls represent Si and Ti atoms, respectively. (b) The local and magnified view of the adsorption site with the isosurfaces of the total charge density showing bonding configuration. (c) Energy landscape. The path of migration of Ti adatom with minimum energy barrier of $E_B$=0.22 eV is shown by stars. (d) Spin polarized energy band structure of silicene including a single Ti adatom adsorbed to each (4x4) supercell. Spin up and spin down bands are shown by blue(dark) and green(light) lines. (e) The corresponding spin polarized total density states and the density of states DOS projected to the Ti atom. (f) Spin polarized energy band structure for a single Ti adsorbed to each (2x2) silicene supercell showing the half-metallic state with a metallic spin up band and a gap between spin down bands.}
\label{fig6}
\end{figure*}

\subsection{Binding of O$_2$ molecule and O adatom}
The interaction of O atom with graphene has been an active field of study. Both experimental and theoretical studies have shown that semi-metallic graphene changes to a semiconductor owing to a gap opening upon oxidation. Reversible oxidation-reduction process and the atomic processes thereof have been treated in several studies. \cite{gox1,gox2,gox3,gox4} It is well-known that the oxidation of silicon surfaces and formation of SiO$_2$ film is one of the crucial processes in microelectronics. Therefore, we expect that O-silicene interaction is even more interesting. In fact, the interaction between O$_2$ and silicene surface is rather strong and leads to the dissociation of molecule into two O atoms provided that the O$_2$ molecule is in a close proximity of the active bridge sites of the silicene. The disassociated oxygen atoms are then adsorbed at different sites of silicene. In ~\ref{fig4} the atomic configuration related with dissociation is presented.

Single O atom binds to silicene above the Si-Si bond at B site. The binding energy is calculated as $E_b$=6.16 eV (591kJ/mol). The bond distance between O and nearest Si atom is 1.70 \AA. The Si-Si bond underneath the O adatom is weakened and is elongated from 2.28 \AA~ to 2.33 \AA. The charge is transferred from silicene to O. The Mulliken analysis estimates the charge transfer from silicene to O adatom to be $Q^*$=-0.34 electrons. The bonding configuration of O adatom and charge transfer to O is presented in ~\ref{fig5}(a-b).

Both sides of silicene are equally reactive and can easily be oxidized by O adatoms. However, the penetration of oxygen adsorbed to one side to the other side is crucial. For example, metal surfaces are protected from oxidation efficiently by graphene coating, since the penetrating oxygen adatom from one side to the other side above the metal surface has to overcome a barrier of $\sim$6 eV.\cite{coating1,coating2,coating3} We did the same test for silicene and calculated the energy barrier necessary for the O adatom to pass from the upper to the bottom side. In ~\ref{fig5}(e) the variation of the total energy is shown as the O adatom is forced to penetrate from one side of the silicene to the other side. In the course of penetration, the coordinates of the atoms are fully relaxed. During penetration, one of the Si-Si bonds expands and the oxygen adatom passes through the center of this expanded Si-Si bond. The energy barrier for this penetration process is calculated as 0.28 eV, which is rather small as compared to the barrier in graphene. In the course of penetration, the Si-Si bond underneath O is slightly elongated and O itself is slightly displaced towards the center of hexagon before it arrives at the equilibrium position below silicene. Low energy barrier implies that O adatoms can easily penetrate into bilayer or multilayer silicene to oxidize them.

The ground state of the system consisting of one O atom adsorbed to the $4 \times 4$ supercell is nonmagnetic. Energy bands of this supercell open a direct band gap of 0.21 eV at the $K$-point of the Brillouin zone, as shown in ~\ref{fig5}(f-g). In this respect, gap opening through oxidation, hence transition from semimetallic silicene to semiconductor is reminiscent of graphene, where controlled reduction/oxidation process by external agents, such as charging or perpendicular electric field was exploited for device applications.\cite{gox1,gox2,gox3,gox4}

\subsection{Ti adatom}
Finally, we consider the coverage consisting of a single Ti atom adsorbed to each (4x4) supercell of silicene. Ti atoms is adsorbed above silicene between T- and H-sites which is identified as the TH site. However, the energetically favorable binding site appears to depend on coverage. For example, the binding site switches to the B site for single Ti atom adsorbed uniformly to each (6x6) supercell of silicene. The binding energy is rather strong and is $E_b$=4.14 eV (397kJ/mol). The minimum energy barrier to the migration of Ti adatom is revealed from the calculated energy landscape in ~\ref{fig6}(c) to be $E_B$=0.22 eV. Hence, Ti adatom is mobile at elevated temperatures. However, the formation energy is negative and consequently Ti clustering on silicene can occur at certain circumstances. Upon binding to TH site, 0.16 electrons are transferred from the Ti adatom to the nearest Si atoms. The binding configuration and other relevant properties of Ti+silicene system are presented in ~\ref{fig6}(a-f).

Ti atom is a light transition metal with an open $3d$-shell and attributes spin polarized ground state to Ti+silicene system. For the system under study, where one Ti adatom is adsorbed to each (4x4) supercell, the magnetic moment per supercell is $\mu$=2.0 $\mu_B$.  Accordingly, spin up and spin down bands split. The linearly crossing bands open a band gap and flat $3d$-bands occur in the gap and around the Fermi level. The minimum gap between spin up bands, $E_{g}^{\uparrow \uparrow}$=0.09 eV occurs at $K$-point. The minimum band gap between spin down bands is indirect and $E^{\downarrow \downarrow}_g$=0.15 eV. However, these electronic and magnetic properties depend on the coverage. Further to the coverage of $\Theta$=1/32, we investigated higher Ti coverage by adsorbing four Ti atoms uniformly on each (4x4) supercell actually leading to the uniform coverage of one Ti atom for every (2x2) silicene supercell (i.e. $\Theta$=1/8). This way we were able to treated also the antiferromagnetic interaction among Ti adatoms. We found that the ground state for the coverage $\Theta$=1/8 is ferromagnetic. Moreover, the gap between spin up bands diminish to make the system metallic for spin up electrons, while the gap between the spin-down bands is reduced. Accordingly, the $\Theta$=1/8 coverage attained a half-metallic behavior.

\section{Conclusions}
The Interaction of Si, C, H, O and Ti adatoms and H$_2$, H$_2$O and O$_2$ molecules are crucial for silicene. Using density functional theory, we examined the energetics of binding and atomic configuration related with these adatoms and molecules. It is predicted that while H$_2$O is non-bonding and H$_2$ is very weakly bound, O$_2$ molecule is dissociated on silicene leading to its oxidation.  Owing to small energy barrier, an oxygen atom bound to the surface of multilayer silicene can easily diffuse to oxidize other layers. We found that oxidized silicene is a semiconductor; the band gap can be tuned by oxygen coverage. Silicene changes to a spin-polarized state upon the adsorption of hydrogen atom. Similar to graphane, the magnetization and the band gap can be tuned by H adatom concentration. Titanium adatoms also attribute coverage dependent magnetic ground state and band gap, which is tuned by Ti concentration. The nonmagnetic silicene becomes a half-metal upon the uniform Ti coverage at $\Theta$=1/8. Even if C adatom forms strong bridge bonding with Si-Si bonds, the substitution of host silicene with carbon adatom is favored. Finally, we revealed that the formation of dumbbell structure by Si adatom can lead to stable structures with interesting coverage dependent physical properties. We believe that the allotropes of silicene consisting of uniform coverage of Si dumbbells will attract interest. It is also shown that silicene acquire diverse and important functionalities owing to its decoration with all these adatoms.

\section{Acknowledgement}
The computational resources have been provided by TUBITAK ULAKBIM, High Performance and Grid Computing Center (TR-Grid e-Infrastructure) and UYBHM at Istanbul Technical University through Grant No. 2-024-2007. This work was supported partially by the Academy of Sciences of Turkey(TUBA) and TUBITAK. The authors acknowledge the financial support of TUBA.

\bibliography{si_func_arxiv.bbl}

\end{document}